\documentclass[sigconf]{acmart}
\AtBeginDocument{%
  }

\setcopyright{acmlicensed}
\copyrightyear{2018}
\acmYear{2018}
\acmDOI{XXXXXXX.XXXXXXX}

\acmConference[Conference acronym 'XX]{Make sure to enter the correct
  conference title from your rights confirmation emai}{June 03--05,
  2018}{Woodstock, NY}
\acmISBN{978-1-4503-XXXX-X/18/06}

\definecolor{darkgreen}{rgb}{0,0.5,0}
\definecolor{orange}{rgb}{1,0.5,0}
\definecolor{teal}{rgb}{0,0.5,0.5}
\definecolor{darkpurple}{rgb}{0.5, 0, 0.5}
\definecolor{olive}{rgb}{0.6,0.6,0}

\copyrightyear{2025}
\acmYear{2025}
\setcopyright{rightsretained}
\acmConference[CHI EA '25]{Extended Abstracts of the CHI Conference on Human Factors in Computing Systems}{April 26-May 1, 2025}{Yokohama, Japan}
\acmBooktitle{Extended Abstracts of the CHI Conference on Human Factors in Computing Systems (CHI EA '25), April 26-May 1, 2025, Yokohama, Japan}\acmDOI{10.1145/3706599.3706676}
\acmISBN{979-8-4007-1395-8/25/04}

\begin{document}

% \title{Prompting More Writing by Speaking: A Diary Study of LLM-Powered Speech Writing Tool in Everyday Context}
\title{Rambler in the Wild: A Diary Study of LLM-Assisted Writing With Speech}

\author{Xuyu Yang}
\authornote{Both authors contributed equally to this research.}
\affiliation{%
  \institution{School of Creative Media, \\ City University of Hong Kong}
  \city{Hong Kong}
  \country{China}
}
\email{xuyuyang2-c@my.cityu.edu.hk}

\author{Wengxi Li}
\authornotemark[1]
\affiliation{%
  \institution{School of Creative Media,\\ City University of Hong Kong}
  \city{Hong Kong}
  \country{China}}
\email{wengxili@cityu.edu.hk}

\author{Matthew G. Lee}
\affiliation{%
  \institution{Stanford University}
  \city{Stanford}
  \state{CA}
  \country{USA}}
\email{mattglee@stanford.edu}

\author{Zhuoyang Li}
\affiliation{%
  \institution{School of Creative Media, \\ City University of Hong Kong}
  \city{Hong Kong}
  \country{China}
}
\email{zhuoyanli4@cityu.edu.hk}

\author{J.D. Zamfirescu-Pereira}
\affiliation{%
 \institution{UC Berkeley}
 \city{Berkeley}
 \state{CA}
 \country{USA}}
\email{zamfi@berkeley.edu}

\author{Can Liu}
\authornote{Corresponding Author}
\affiliation{%
  \institution{School of Creative Media, \\ City University of Hong Kong}
  \city{Hong Kong}
  \country{China}}
\email{canliu@cityu.edu.hk}

% \affiliation{%

% \affiliation{%

% \affiliation{%

\renewcommand{\shortauthors}{Yang, Li, et al.}

\begin{abstract}
  % Traditional writing is notoriously difficult to start, but 
  Speech-to-text technologies have been shown to improve text input efficiency and potentially lower the barriers to writing. %for writing people writing faster, with the assistance of large language models. 
  Recent LLM-assisted dictation tools aim to support writing with speech by bridging the gaps between speaking and traditional writing.  
  This case study reports on the real-world writing experiences of twelve academic or creative writers using one such tool, Rambler, to write various pieces such as blog posts, diaries, screenplays, notes, or fictional stories, etc.
  %creative writers (e.g. academic faculties, Ph.D. students, bloggers) writing articles by speech and editing with LLM features.
  Through a ten-day diary study, we identified the participants' in-context writing strategies using Rambler, such as how they expanded from an outline or organized their loose thoughts for different writing goals. %that they write in real-life settings by expanding from the dictated outline or organizing the free-speaking content. 
  The interviews uncovered the psychological and productivity affordances of writing with speech, pointing to future directions of designing for this writing modality and the utilization of AI support. % This writing experience can enhance natural and emotional expression, boost productive thinking, and foster writing self-efficacy. Yet, it also introduced new challenges and provided important implications about how to design future AI-based tools for writing with speech.
\end{abstract}


% \keywords{Do, Not, Us, This, Code, Put, the, Correct, Terms, for,


\maketitle
\section{Introduction}
In the realm of writing tools, the integration of speech input technology has emerged as a transformative force, offering users an alternative mode of interaction that transcends traditional paper \& pen-based or keyboard-based methods~\cite{lin2024rambler, Ruan2018Comparing}. Leveraging speech input to empower writing stems from speech's inherent ability to bridge the gap between thought and text, allowing users to articulate their ideas naturally and fluidly~\cite{lin2024rambler, Mehra2023GistAV, Crystal2005SpeakingOW, Ruan2018Comparing}. Through speech, individuals can bypass the constraints imposed by manual typing, enabling a more direct and expressive form of communication~\cite{lin2024rambler}. However, challenges such as ambient noise interference, dialectal variations, and inaccuracies in speech recognition systems can hinder the accuracy and dependability of transcriptions, necessitating additional efforts to refine the converted speech text~\cite{Karat1999PatternsOE}.

Previous research has explored the use of natural language processing (NLP) and large language models (LLMs) to assist in writing~\cite{Dang2022BeyondTG} and reviewing spoken dialogue~\cite{Li2021HierarchicalSF, Li2023ImprovingAS}, especially in cleaning disfluency and speech recognition errors~\cite{Liao2020ImprovingRF, bassi2023end, Tanaka2018NeuralEC}. Users can also shorten, summarize or replace the selected text with LLM suggestions or prompt the model for text generation~\cite{Yang2022AIAA}. In addition, LLMs can support macro-level structural revision, which moves beyond automatic text summarization to semantic manipulation with writers in semantic control~\cite{Arnold2021GenerativeMC}. For example, Dang et al~\cite{Dang2022BeyondTG} introduces a writing tool that provides on-the-fly paragraph summarization along with the original text. Users could interact with the summaries of paragraphs, such as reorganization via drag and drop, which manipulates the original text in parallel.

Our recent research~\cite{lin2024rambler} presents an LLM-powered graphical user interface, called Rambler, that supports gist-level manipulation of dictated text with two main sets of functions: gist extraction and macro revision. To evaluate the effectiveness of this approach, we conducted a lab study, where 12 participants were asked to compose two articles each, one using Rambler and the other using a baseline (a standard STT editor + ChatGPT). Participants chose from broad writing topics provided by the experimenter for completing the tasks. The findings showed an overwhelming preference for Rambler over the baseline (10 out of 12) and demonstrated diverse writing strategies adopted by participants. However, while a lab study in a controlled environment ensures comparability between conditions, it lacks real-world context, and could not test the middle- or long-term usage of the approach. In our case, since participants were writing content for a lab study, there was low incentive for them to edit it as thoroughly as they might for their own writing in real life. Additionally, there might not be enough time in a one-hour lab study for participants to adopt a new writing paradigm using speech as text input modality. 

As a follow-up work, we conducted an in-the-wild study to understand the real-world longer-term usage of Rambler for creative writing tasks. The goal was to understand the affordance of writing with speech and its user acceptance as well as how the functions of Rambler could be appropriated into users' diverse writing habits.  

Through a few focus groups and a ten-day diary study, participants wrote three articles on their own topics with personal device(s) and in their real-life environment. We report on participants' writing strategies and workflows led by distinctive writing goals. The surveys and interviews about their experiences provide insights into their incentives and obstacles in adopting the speech modality for writing and the role of AI assistance in the process. 

Based on the findings, we discuss the implications of designing STT-based writing tools and opportunities for AI support.
\section{Background of a Speech-Based Writing Tool -- Rambler}

\begin{figure}
  \Description{A labeled screenshot of the Rambler UI. Ramble in default state, with revision functions accessible through buttons on the top of each paragraph, a paragraph is called Ramble. The first Ramble shows re-speaking mode, where voice input is transcribed so that it can be appended to current text, replace the current text, or to be discarded using the buttons on the right corner at the bottom of Ramble. The buttons at the bottom of the UI: the Semantic Merge button, New Ramble button, and Semantic Zoom slider.}
  \centering
  \includegraphics[width=0.95\linewidth]{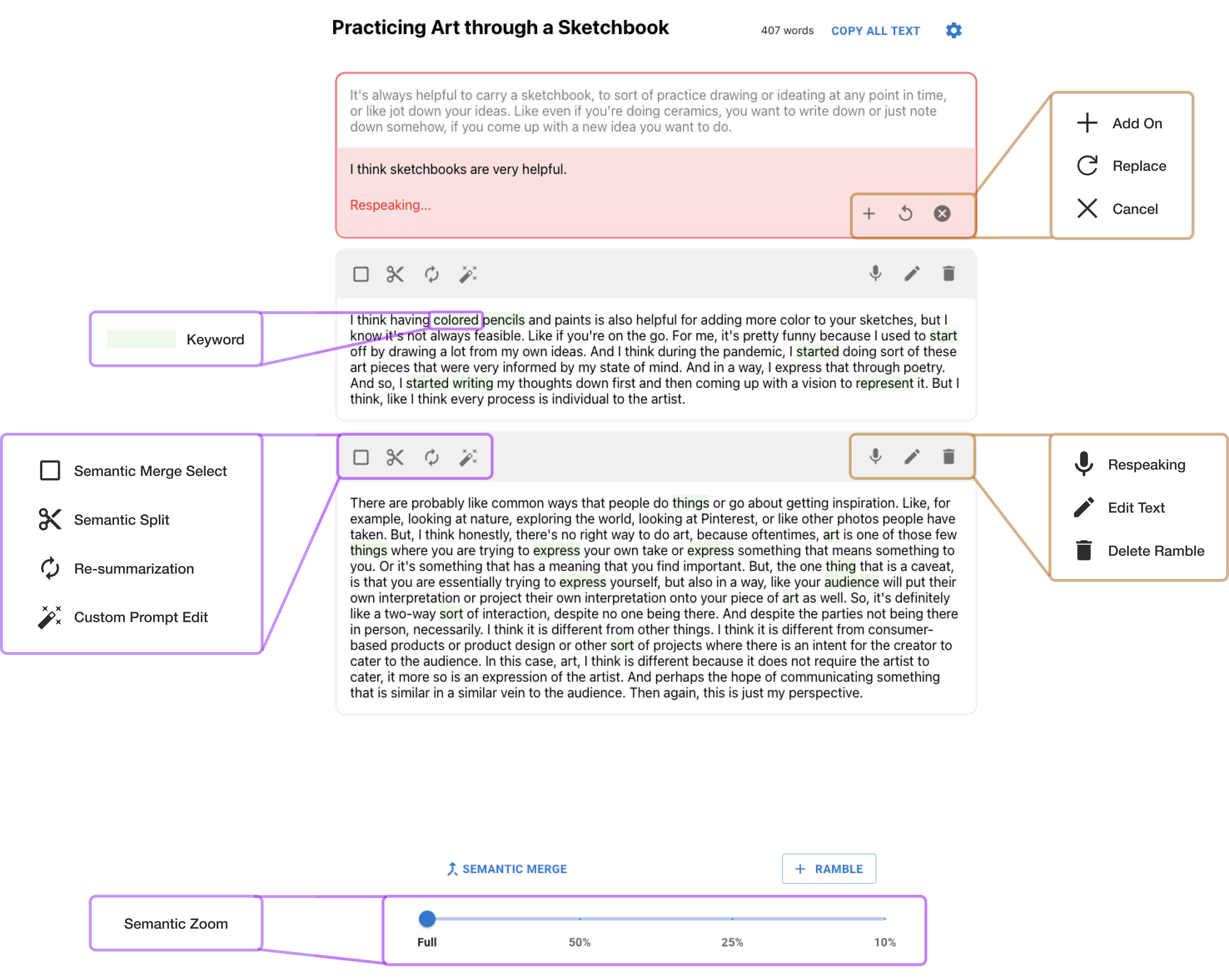}
  \caption{Rambler interface on a tablet from~\cite{lin2024rambler}. Users can dictate content into individual Rambles and use various semantic and manual functions to macro-edit and reorganize them.}
  \label{fig:rambler}
\end{figure}

The speech-to-text tool we used is Rambler~\cite{lin2024rambler}, which has users dictate their impromptu thoughts, placing each segment of recording into a \textit{Ramble}. Rambler provides three main functions. First, it supports users in dictation by cleaning up transcripts. Each recording is transcribed by a real-time STT API and then preprocessed with an LLM that automatically cleans up any disfluencies and punctuation errors, completes broken sentences, and smooths transitions in the raw transcript. Users can also use the \textit{Respeaking} function to dictate new content to replace or add to the existing Ramble (the microphone button in Figure~\ref{fig:rambler}). Second, it can extract the \textit{gists} in the text to support visualization and interaction with that text. The keywords of the content are shown automatically after recording, and users can adjust the selection on demand. Based on the gists selection, Rambler provides \textit{Semantic Zoom} to show multiple summarization levels of each Ramble (the bottom slider).

Besides, Rambler provides several LLM-based functions for users to perform macro-revisions on the content. \textit{Semantic Split} operates on an individual Ramble to divide one Ramble into N Rambles based on its content (the scissors button in Figure~\ref{fig:rambler}). \textit{Semantic Merge} merges the contents of multiple selected Rambles into one Ramble (the merge button in Figure~\ref{fig:rambler}). \textit{Custom Magic Prompt} opens up the possibility for users to define any custom transformation on Ramble content by directly inputting an LLM prompt (the magic wand button in Figure~\ref{fig:rambler}). Rambler also provides manual editing functions. \textit{Manual Merge} concatenates the text of two Rambles into one Ramble (through drag and drop). \textit{Manual Split} splits a Ramble into two at the cursor position. Users can always drag and drop the Ramble to reorder or edit the content using the keyboard. 

\section{Methods}

The goal of the case study is to understand how writers use speech as the primary text input modality for their own writing tasks in a real-life context, as well as how AI features could effectively support them. 
Our research questions are: 1) How do academic/creative writers use a speech-based writing tool to write in real life? 2) What new affordances does the speech modality bring and how could AI support writing with speech? 3) What factors affect the user acceptance of this new writing paradigm?

We began with a few focus group sessions gathering the recruited participants to discuss their writing habits, get introduced to the Rambler tool, and brainstorm their envisioned scenarios for using it. Subsequently, participants could opt in for a ten-day diary study where they would write a minimum of three articles. The study involves three intermediate surveys and one exit interview upon completion.

All procedures were approved by the university’s research ethics review (Human Research) before data collection.

\subsection{Participant}

A total of 14 respondents registered to participate in the focus groups.
The selection of participants in the focus groups was tailored to involve a diverse set of individuals interested in various types of creative writing, with experience or curiosity about dictation. We did not have any specific style or literary genre prerequisites. All participants were not English native speakers, but participants had to qualify for the case study through either experience in a English-taught degree program, a score over 90 in TOEFL (Test of English as a Foreign Language), or a score over 6.5 in IELTS (International English Language Testing System): their English proficiency information was gathered through surveys.

For the subsequent diary study phase, a total of 12 participants were involved. 11 out of the 14 initial focus group participants opted to join, with an additional participant (P15) who missed the focus group sessions due to logistic reasons.
Detailed participant information is in Appendix Table~\ref{tab:users}.

\subsection{Study Design and Procedure}

\subsubsection{Focus Group}
The purpose of the focus group was to make sure that the participants got familiarized with the research tool Rambler and actively thought about how to utilize this new way of writing. Four rounds of focus group sessions were held, three were online and one was on-site, grouped based on the participants' availability. Each focus group was conducted in three steps. In step 1, participants were asked to share their current writing habits, strategies, and experiences. In step 2, the features of Rambler were introduced to the participants via a video overview and live demonstration with Q\&A. Participants were given a few minutes to try out the functions. In step 3, the host initiated a brainstorming for participants to envision their potential writing tasks and scenarios using Rambler to serve their own writing goals in real life. Each focus group session lasted about one hour. The participation of a focus group session granted each participant a 100 HKD supermarket coupon as compensation. Interested participants were asked to register for the diary study at the end of the session. 

\subsubsection{Diary Study}
After each focus group session, registered participants began their diary study phase, which lasted 7 to 10 days. The registration survey asked about the participants' demographic information, writing topic interests, dictation experiences, and LLM experiences. They were asked to use Rambler to write throughout the study period and finish at least 3 articles, each of at least 500 words, by the end of the study.
Participants were allowed to freely choose their writing tasks and purposes but were required to finalize the draft for their chosen purposes. As Rambler is a web application, the participants could use it via a web browser on any device via a unique URL generated for each participant. They were warned to avoid any confidential content as the content is logged without encryption. 
Every two days, they were required to fill out a survey sampling their usage situations (writing content, devices of use, preferences of functions, and any insights or problems discovered during their usage). During the diary study, we sent a reminder every two days, either by text or email based on preference, to remind participants to perform the task 
and fill out the experience survey on time.

\subsubsection{Post-study Interview}
Interviews took place online or on-site per participants' requests. The questions in the interviews were geared toward study goals, including the affordance of writing with speech, comparison to existing writing methods, user strategies from ideation to creation processes, user acceptance, and suggestions. Upon completion, compensation for the diary study was a HKD 400 supermarket coupon per participant. Each interview took one hour.

\subsection{Data Collection and Analysis}
We collected screen recordings of the focus groups via an online conference tool Zoom. For the diary study, we collected thirty-six articles in total (each participant completed three articles, with each one at least 500 words), audio recordings of the exit interview, and three surveys from each participant. Participants' interactions with the tool were logged in the Rambler application on our web server. We programmatically processed the logs to analyze how content evolved through the use of Rambler features.
We collected qualitative data from the audio transcriptions of the focus groups and interviews and the written answers from the diary surveys. Inductive thematic analysis was used to identify patterns and themes within data~\cite{fereday2006demonstrating}. Two researchers coded 25\% of the data independently and discussed their interpretations to reach a consensus. Then one of them coded the rest of the data. Multiple researchers worked on the categorization of codes into themes and patterns together. We iteratively reviewed and refined the identified themes regarding users' strategies, affordances, and user acceptance of LLM-powered speech-based writing, to ensure they accurately reflected the data and provided a comprehensive representation.
\section{Findings}
We report qualitative findings here from the focus group and diary study interviews and surveys. We first draw an overview of the participants' existing writing habits and envisioned use of Rambler from the focus group, then answer each research question with themes identified from thematic analysis. We end this section with a list of design suggestions made by participants.
\subsection{Focus Group - Participants' Envisioned Usage Scenarios}
\subsubsection{Participants' Writing Habits}
Participants' motivations for writing varied from emotionally significant moments, unique experiences, to academic or work-related writings. They talked about their writing on mobile phones, which was handy for capturing random thoughts (P3, P5, P6). They liked to utilize specialized tools to aid in various writing stages such as idea storage (P8), topic management (P5), logical restructuring (P5), and grammar refinement (P9). They also explored tools to effectively segment and organize multiple writing projects (P2).

Some participants had experience with dictation and mentioned their use of it to multitask or avoid typing (P1). There was a perceived text input efficiency (P5) and benefit in language practice (P6). Some found capturing spoken content disruptive or awkward (P2, P3, P4), with concerns regarding accuracy and responsiveness in noisy environments (P10). Participants who were familiar with LLM tools heavily relied on them for various tasks, including ideation (P1, P2), providing initial ideas (P1), finding words (P2), revising tone (P4), improving grammar (P9), generating specific writing characteristics (P2), learning concepts (P1), practicing writing (P8), improving skills (P6), and translating text (P6).

\subsubsection{Envisioned Use Cases and Scenarios of Rambler}

The participants envisioned using Rambler to assist with schoolwork (P5), learning skills (P6), capturing conversation information (P2), generating functional content (such as emails) (P5), diary writing (P4, P6), screenplay creation (P11, P12), and scripting episodes for podcasts (P10).
Moreover, they liked the flexibility of using Rambler on any device, and envisioned utilizing it while walking on the road (P7), lounging in bed (P2, P4, P6), or relaxing outdoors in a park (P10). This flexibility in usage locations improves the acceptance of Rambler, catering to users' diverse writing preferences and habits.

\subsection{How do creative/academic writers use a speech-based writing tool in real life? }
In the subsequent diary study, participants generally performed their envisioned writing scenarios except for a few topic changes due to technical or logistic constraints. Table~\ref{tab:users} in the Appendix summarized what each participant wrote, in what environment, and on what device. We can see that the majority of writing tasks were done on mobile phones (56.8\%), primarily on a desk or in bed with only a few occasions during walking. Some used the computer (43.2\%) for a larger display and convenient operation. The surveys asked how they distributed their time for writing. The answers showed that 42\% of the writing tasks were completed in one go, while the rest of the tasks were done in a ``distributed'' manner. One task was done after several sessions, once per day or half a day based on time or location in their routine. 

Like the lab study findings of Rambler~\cite{lin2024rambler}, in the diary study we also observed the two distinct writing strategies: 1) \textit{outline first}---users create an outline through dictation and then expand on it; and 2) \textit{free-speaking}---users dictate detailed content spontaneously before editing it. The key difference between these two patterns lies in whether users formulate their narratives before or after dictation. From the logged content, we noticed that the former strategy tended to be used in academic or communicative writings with an external audience, while the latter strategy was more used in personal or reflective writings for oneself. 

\subsubsection{Outline Expansion for Academic or Communicative Writing.}

\begin{figure}
    \Description{This figure illustrates a structured workflow for content development and editing, where ideas progress from outlining to expansion, refinement, reordering, and manual merging to form a cohesive final document.}
    \centering
    \includegraphics[width=0.95\linewidth]{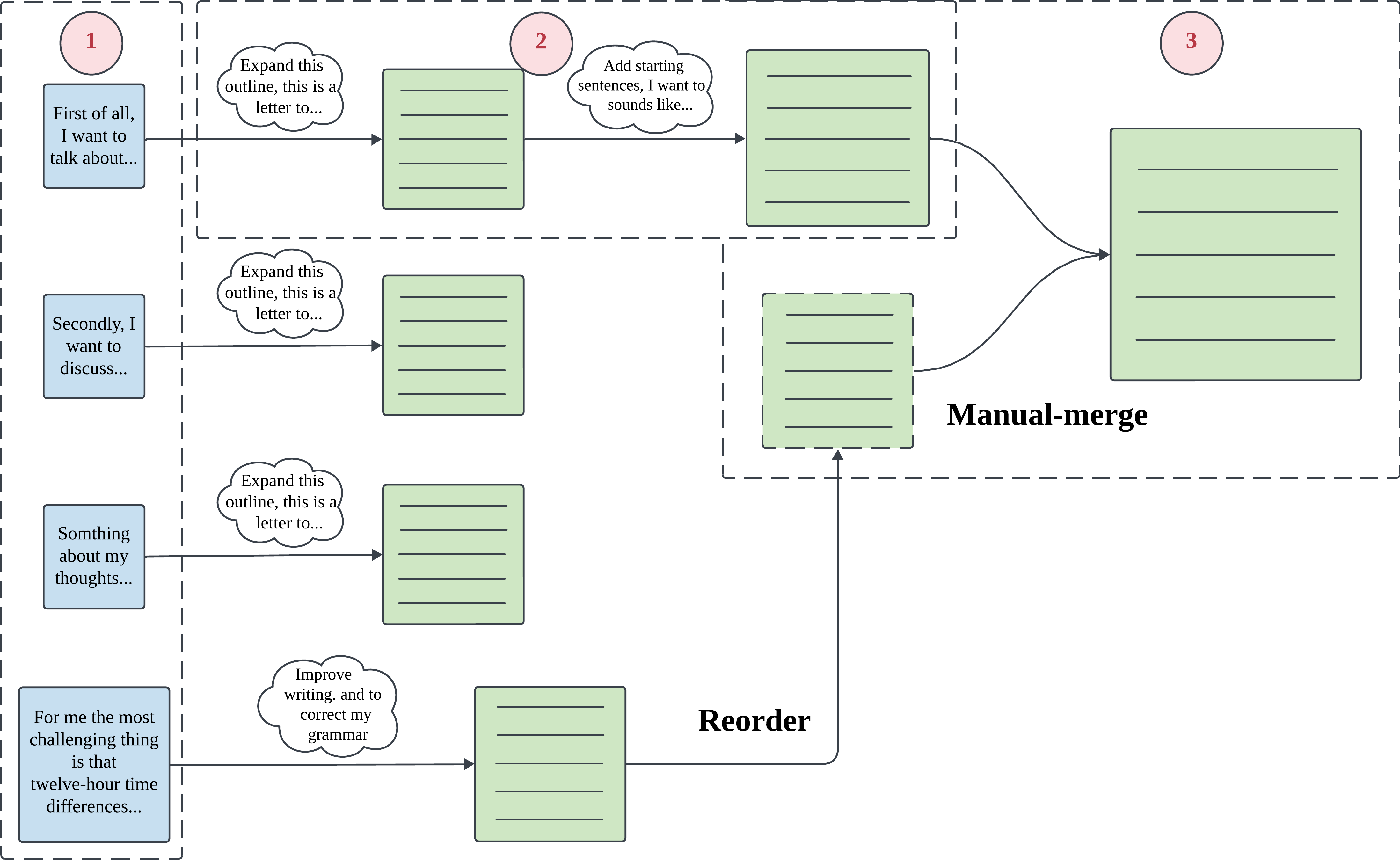}
    \caption{Participant P2 wrote a letter by dictating an outline first and expanding from it. The numbers represent the order of actions. Blue ones are generated by dictation, and green ones are generated by macro revision.}
    \label{fig:str1}
\end{figure}

When participants have clear ideas about what they are going to write, they choose to dictate key points into each Ramble and expand those points with AI-powered semantic functions. Such writing patterns often occur when participants craft logically structured writings and have envisioned the structure, though not necessarily the details, before writing (P2, P3, P9, P14). For example, when writing a responding letter, P2 began by dictating five paragraphs that highlighted different key points to create an outline (see Figure~\ref{fig:str1}): they mention ``\textit{this is a letter I want to respond to his previous letter so I already have the structure in my mind which point I want to reply.}'' After creating the outline, each Ramble was processed through a custom magic prompt to expand the content, such as \textit{``expand this outline to a full-text paragraph''} and \textit{``add an example at the beginning or the end.''} By doing this step, the LLM adds more detail to the outline, and the length of each paragraph goes from one sentence to about five sentences. P2 reflected on this experience, stating, \textit{``I usually don't have enough time to reorganize that into a like complete article. But every letter I sent to my boyfriend is like complete article, so I think it's content I can use to finish the task.''}

Aside from relying on AI-generated content, some participants enhanced their writing by dictating more to supplement the content added by Magic Prompt or by re-organizing the content inspired by AI-generated content.
For instance, when P3 prepared a presentation script to update instructors about a project, they began by outlining their ideas based on existing slides. They then used Magic Prompt to refine the script. In the process, the AI-generated content often sparked memories of additional details, enabling P3 to elaborate on their initial outline. As P3 noted,\textit{``I adjusted the slides in the process, correspondingly, I also adjusted the content of my scripts.''}
It is common that when participants want to add some content to the end of a Ramble, they usually create a new Ramble and merge it with the previous one instead of respeaking to add content.

\subsubsection{Capturing Loose Thoughts in Personal and Reflective Writing.}

\begin{figure}
    \Description{The figure illustrates how larger ideas are broken down into focused segments and later recombined for improved organization and tone adjustments.}
    \centering
    \includegraphics[width=0.95\linewidth]{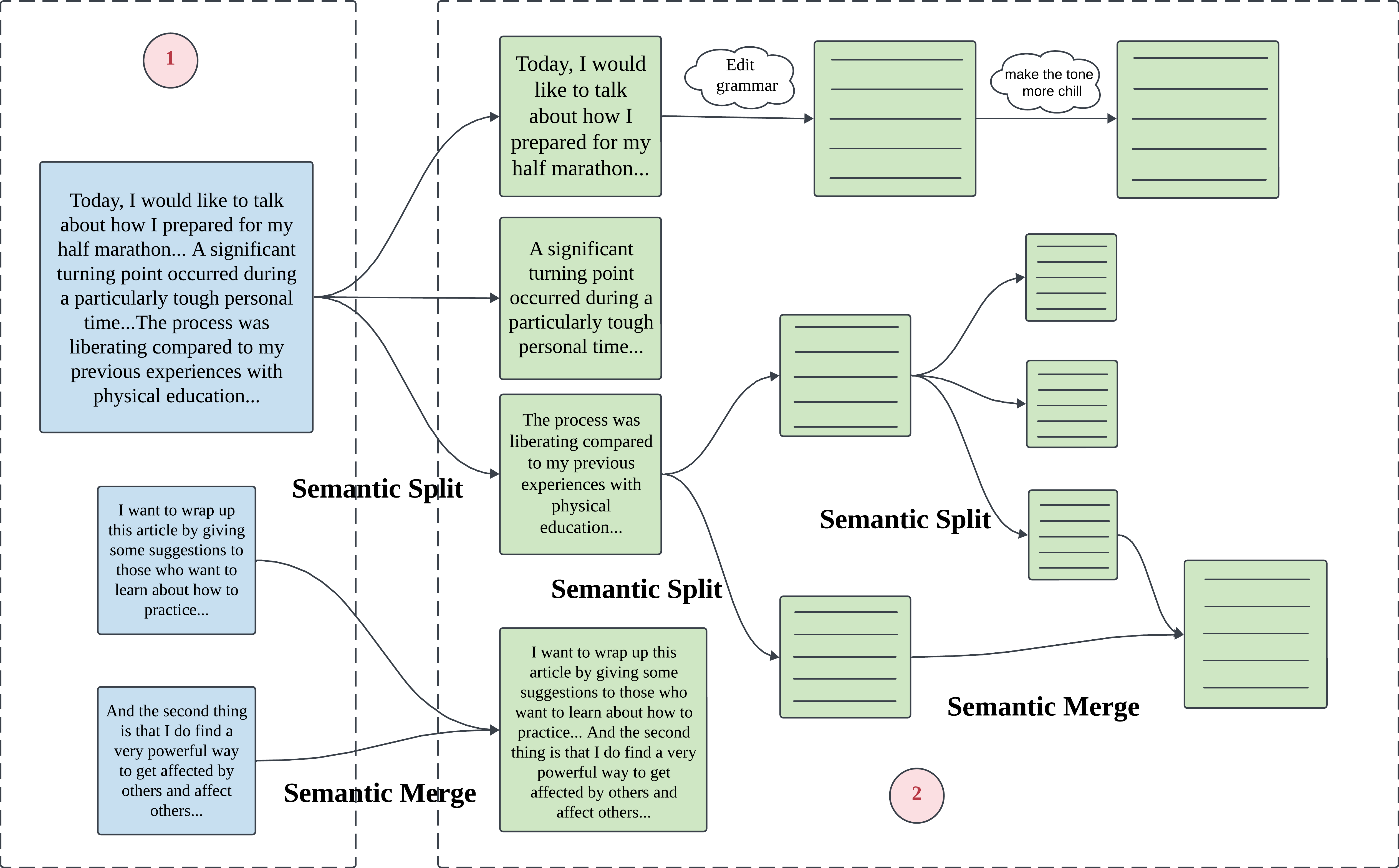}
    \caption{Participant P4 wrote an experience sharing by speaking detailed content and reorganizing it. The numbers represent the order of actions. Blue ones are generated by dictation, and green ones are generated by macro revision.}
    \label{fig:str2}
\end{figure}

As P5 said, \textit{``I usually spoke a big chunk to Rambler at the beginning.''} Instead of dictating in a typing-like way, more articles were created by speaking everything in one go at the beginning, especially when participants knew the details to be described, such as writing diaries (P4), emotional notes (P6), and comments about a movie (P14). As shown in Figure~\ref{fig:str2}, P4 depicted the day's itinerary in detail. Then, the Ramble with a long paragraph is split into three short Rambles by Semantic Split. Each Ramble marks an event in the diary. For each Ramble, P4 used Magic Prompt to improve grammar, add more details, and change the tone to be more chill and easy to understand. Finally, P4 used Semantic Merge to combine two related Rambles to finish the draft.

Since users do not have a predefined outline when writing, subsequent restructuring is more common than the previous strategies. P5 reports that Semantic Split and Merge can meet their expectations in restructuring: \textit{``Semantic Split made the initial chunk paragraph into a few smaller paragraphs, and the qualities of the smaller paragraphs normally met my satisfaction.''} P5 also found that Semantic Split helped create paragraphs that started with key points, clearing up long paragraphs: \textit{``It is helpful for breaking text into manageable paragraphs with clear topic sentences''}. P12 believed that Semantic Merge helped merge similar paragraphs and encouraged them to rethink the structure so that they had a clearer direction for writing: \textit{``Semantic Merge is the main reason why I use Rambler because I need to sort out my thoughts in a logical way when brainstorming, rather than simply placing words with STT, which doesn’t help me save time.''}

After reorganization, Magic Prompt is widely used to adjust the writing style and tone. For example, P5 asks Rambler to create formal, grammatically correct writing by eliminating pauses and expressions while preserving the original meaning: \textit{``Improve writing to be formal. and correct my grammar but keep my meaning''}. To improve the performance of Magic Prompt, users gave several demands in one prompt and provide the context: \textit{``Make the tone sounds more chill and use words easy to understand it’s my diary don’t be too serious.''} (P2) When they are not satisfied with the results, they will iterate on the prompts until satisfaction, such as \textit{``make it more intense''} and \textit{``Not that intense. Make it sound like he is scared. But more anxious.''} (P8) Aside from macro-revisions, manual editing is generally used to polish the wording, such as short phrases or punctuation: \textit{``When I knew the exact word(s) I wanted to add, I would add it by typing.''} (P9)

\subsection{What new affordances does writing with speech have and how could AI help?}

We summarized the following findings about the affordances of writing with speech, as a new way of writing that people perceive and feel in their experience of using Rambler, including how it facilitated emotional expression, boosted writing productivity, and improved their self-efficacy. 

\subsubsection{Speaking as a Natural and Emotional Expression Channel and a Communicative Act.}
The inherent quality of using speech mirrors human natural conversation, thus offering users an uninterrupted channel for authentic emotional recording and introspection. 
P3 likened the nature of writing with speech to an undisturbed telephone conversation:\textit{ ``Doing speech-based writing is like talking on the phone and would not be interfered with.''} 
P5 highlighted the emotional resonance enabled by dictation: using speech to write about emotional events would capture the emotions so vividly and authentically for them that upon reviewing the transcript, they could feel how they felt at that moment.
This unfiltered self-expression makes speech-based writing suitable for personal diary. 
As P2 explained, \textit{``I could be the real me when using dictation to write a diary.''}
P6 noted, \textit{``It's just like you speak to another person, but that person actually doesn't exist and you can see whatever you want to see to it.''}

Furthermore, the act of expressing emotions through dictation serves as a vehicle for organizing one's psychological landscape and fostering introspection, like a meditative experience.
As P4 describes \textit{``using Rambler feels like meditation or self-reflection''}.
Speaking is also associated with collaborating with a teammate, facilitating communicative production.
P4 highlighted that writing with speech was like chatting with a close person, expanding the branches of thoughts together. P13 also noted that the process of writing with speech was like talking with someone by phone, whose purpose was expressing their thoughts.
Considering this property, one participant decided to use Rambler to write a screenplay and found it highly satisfactory. P12 said, \textit{``Rambler directly captured my very simple words, the output generated by dictation surprised me, since the format and literary style was very close to screenplay. In screenplay, you don't need to write in a sophisticated way. So I realized that a screenplay can be simply completed, and I just finished this one (screenplay) in a very cozy sitting posture on the sofa.''}

\vspace{-2mm}
\subsubsection{Speaking Boosts Productivity and LLM Prevents Overthinking.}

The efficiency of dictation
often lead to a surprising experience with the substantial volume of text generated by talking. P14 was notably impressed by the productivity of using dictation
after merely speaking for 1 to 2 minutes, which produced a few hundred words.
They felt it lowered the pressure of writing tasks. P8 echoed a similar experience after a few days of usage and stated that it became easier to output several hundreds of words in one go compared to typing every single sentence on the keyboard.

Sometimes writers overthink a detail and get stuck as they fixate on word choice from the beginning. Perfectionism can block individuals’ inspiration and hinder their creative thinking. Using Rambler, participants felt that the LLM features nudged them to wisely assign their attention and energy towards deliberating the overall content and structure of their work.
P12 noted that they could save energy to think about new ideas with the assistance of LLMs in crafting the wording details. The fluent experience of ideation with the aid of LLM helped prevent individuals from overthinking and reduced their perfectionism in the writing process.
As P8 stated, \textit{``Writing with speech is beneficial, especially with preventing me from overthinking sentences, helping me go with the flow, and leaving some of the more complex sentence building to AI.''}

\vspace{2mm}
\subsubsection{LLM Features Foster Self-efficacy.}~\label{efficacy}
The LLM features for polishing the oral spoken draft into written formats brought confidence to users in their writing process. P4 commented that LLMs could assist formal writing in good quality since it was able to rephrase their spoken tone to a suitable style.
P14 also expressed appreciation for this aspect: \textit{``Since LLM could polish the wording in a beautiful style, I didn't need to worry about whether my speaking was correct or not, instead, it helped to remove the errors in my original speaking at the beginning, it effectively avoided the errors.''}

Moreover,
participants prompted LLMs to adopt specific literary styles, and improved their language learning by observing the polished outcomes. 
P14 noted personal growth in language proficiency through this feature. As the LLM polished their originally simple wording into a higher quality, they learned new writing styles and skills from the change and realized that it could be a good opportunity to improve English writing skills.
P2 was amazed after employing an original English diary writing style by prompting the LLM:
\textit{``Rambler helped me to make my diary look more like an English native speaker, which impressed me a lot. I felt surprised by the change since it is highly like the style in a book about interesting English diaries I read before, it's a way to enhance my writing skill and make my writing close to the original English literary style.''}

\subsection{What factors affect the user acceptance of writing with speech?}
\begin{figure}
    \Description{A boxplot illustrating how participants felt comfortable during three rounds of writing tasks (1st, 2nd, and 3rd).}
    \centering
    \includegraphics[width=0.9\linewidth]{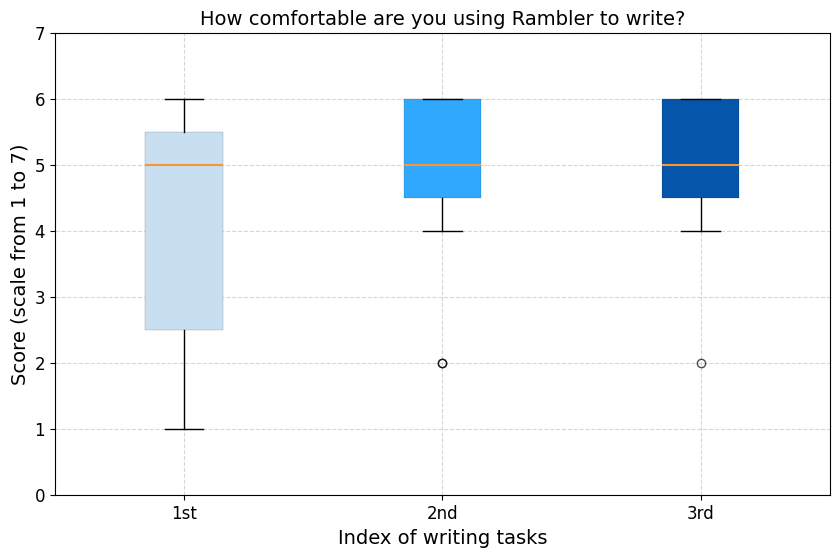}
    \caption{Participants' user acceptance during three rounds of writing tasks using a 7-point Likert scale.}
    \label{fig:useracc}
\end{figure}

Nine out of twelve participants (P2, P3, P4, P5, P6, P9, P13, P14, P15) reported that they were able to get used to writing with speech in the short period of the diary study, while three (P8, P10, P12) expressed reservation. As shown in Figure~\ref{fig:useracc}, the median of the score for ``comfort of use'' is 5 (out of 7) for all three tasks submitted over time, while the deviation of the scores decreases after the first use. This shows positive user acceptance of this new approach for writing, as well as some initial learning curve to get used to it.

\subsubsection{Productivity Gain.} Although we could not measure task completion time effectively in this study due to its in-the-wild nature, we did ask for an estimation of time they spent on a writing task in each survey. Their answers showed, 66.7\% of the 500-word writing tasks were completed in 10 to 30 minutes, 15.4\% of them needed 30 to 60 minutes, 10.3\% over an hour, 7.7\% within 10 minutes.
Participants felt a sense of achievement after completing a task with Rambler and expressed positive surprises in gaining trust in the technology. For P14, because writing was a heavy task that brought a huge mental burden, it was previously hard for them to complete an article in traditional ways (either typing or handwriting), and so they gave up having a writing routine.
In the focus group discussion, they initially expressed doubts about whether speech could facilitate writing and thought dictation technology employed in writing was an ambiguous concept to them. Yet in the post-study interview, they shared a big sense of surprise from the productivity gain in writing with speech.
They expressed willingness to restart a writing routine in the future via dictation. 

\subsubsection{Effective help for organizing thoughts.} Another reason for participants to adopt this technology was that the LLM features could greatly reduce their effort by merging several vague ideas into one. P12 said in the focus group that using traditional writing to illustrate details of ideas from a vague concept could always block their writing. 
In the post-study interview, they shared their pleasant experiences of inputting their vague ideas and prompting the LLM to merge them into a coherent paragraph. They felt that the LLM features could effectively and efficiently boost clarity at the ideation stage.

\subsubsection{Remaining Challenges.}
In the meantime, there remain some challenges in adopting this new writing paradigm.
One challenge the participants faced was that they felt distracted during the process of speaking while thinking about the next sentence (P8). 
Although Rambler leveraged LLM to correct recognition errors and disfluencies, detailed editing is still necessary at times.
P3 expressed the inconvenience of having to modify a typo in the middle of a sentence. Unlike typing on keyboards, it is nearly impossible to direct the cursor to a certain place via LLM prompting.

Last but not least, several participants raised the issue of being unable to multitask during speech composition. For instance, participants felt that they could not use Rambler to write something that requires them to research online at the same time. P2 said, \textit{``Using Rambler on phone is not convenient for reading and doing research meanwhile.''}
P3 also mentioned the inconvenience of how Rambler currently cannot support multitasking in different tabs or apps.

\subsection{User-Suggested Improvement for Design}
\subsubsection{Context Awareness} Participants made suggestions of reusing their Custom Magic Prompts. As P9 stated, ``\textit{It would be better if Rambler could save my prompts' history, so I could quickly access them again if I needed.}'' P2 said, ``\textit{It would be more useful if the prompt can be applied in all the Rambles simultaneously on the purpose of efficiently knowing the context of my article and adjusting the writing tone of the whole article in one go.}''

\subsubsection{Personalized Conversational Support} Given the nature of speaking as a communicative act, integrating a virtually personalized assistant in writing with speech could represent a strategic enhancement. P14 said, ``\textit{It could encourage users to have more inspirations to write, if there is a chatbot or assistant knowing the context on Rambler, since the behavior of having conversation or communication with them can enhance people to think forward.}''
P2 attempted that prompting an assistant-like suggestion but did not get the expected result, what they did was giving an identity to the LLM, and expected the feedback from LLM to generate more human-like information for personal content. Similarly, P9 had the same feeling as well in the creation process: \textit{“It could be very useful to me if I could personalize a chatbot on Rambler”.}

\subsubsection{Alternative Information Storage and Display} Some participants felt it would be good to keep their original audio mapped to the transcripts for error prevention (P9, P15). Others also wanted to be able to see the text and outline side-by-side so that they could read the main points of each paragraph at a glance during composition (P15). Participants also requested easier ways to distinguish the Rambles, such as by adding a title for each. P9 said, ``\emph{It would be better if there's a title for each Ramble ... I had to re-read them every time if I wanted to remind myself of the main point of each paragraph. The process could be annoying and consumed my patience.}''
\section{Discussions}
\subsection{In Relation to the Lab Study}
The lab study evaluated Rambler against a baseline and observed the usage of individual functions. It was found to be advantageous in helping users review spoken compositions and organize their thoughts, while showing high versatility in supporting diverse writing strategies.
Consistent with these findings, the diary study confirms that Rambler helps users organize thoughts, boosts efficiency in outlining and expanding ideas. In the real-world context, we found it also inspired reflective writing and journaling. While the lab study findings centered around the usage pattern and experiences with individual features of the tool, the diary study focused more on a holistic interpretation of the overall experience with the new writing paradigm. It showed good user acceptance for this approach in real-world creative writing scenarios for the enhanced productivity, the natural and emotional act of speech and its ability to prevent overthinking.
In terms of the usage patterns in the real world, we saw much flexibility in how participants distributed their writing times over days and hours. Perhaps the use of speech input on mobile devices facilitated distributed writing. In addition, the primary writing environments were in quiet places around the desk or in bed, rarely in outdoor places, highlighting the constraints for using speech input due to noisy and distraction.

\subsection{Speech Input and LLM Features go Hand-in-hand}
The findings of our study revealed that using speech as the main modality for writing carries unique advantages, including a large productivity gain and the psychological benefits of emotional or introspective expression. Yet due to these characteristics, the content being produced tends to be abundant. Therefore, the LLM features of semantically extracting information, organizing multiple pieces into a whole, splitting disconnected points and polishing the wordings all go hand in hand with it to complement the shortcomings of spoken production. 

\subsection{Writing for What and Who Matters}
Two distinctive strategies of composition led to two major directions of speech-based writing, one for academic or communicative purposes with external audiences; the other for introspective expression by recording feelings and experiences for self reflection. The usage and preferences of functions differ largely between these two types of writing scenarios: one starts with an outline and utilizes expansion methods and the other starts with loose thoughts and using merging methods to converge. Future work could consider optimizing each direction with specialized tools and information representation methods. 

\subsection{Writing With Speech is a Viable Paradigm} 
This in-the-wild study tested whether users could get used to writing with speech in a short period. Participants' positive feedback showed promise for this approach. Productivity gains and the lowered barrier for writing played a pivotal role in user acceptance, accompanied by the effective clarification of vague ideas through the act of speaking and iterative organization. Future designs could dive further into adapting to users' context and purpose of writing, such as addressing the need to refer to external materials in research-based writing. Context-awareness and personalized conversational support can be the natural next step for a speech-based writing tool. 

\section{Conclusion}
This work evaluated the real-world experiences of using speech-to-text as the primary text input method for writing. By analyzing twelve academic and creative writers' writing experiences with an LLM-assisted dictation tool, we conclude with a positive outlook for this new writing paradigm based on its productivity gain and psychological benefits. Our insights on the effectiveness of features based on different writing purposes pave the way for future generations of AI-supported writing with speech.

\begin{acks}
This research is partially funded by Google Faculty Research Award (CityU Hong Kong 9229068) and the Berkeley Artificial Intelligence Research Lab - Open Research Commons. We thank Susan Lin, Björn Hartmann, Michael Xuelin Huang and Shumin Zhai for their invaluable support. 
\end{acks}

\bibliographystyle{ACM-Reference-Format}
\bibliography{references}


\begin{thebibliography}{14}

%%% the Web address.

\ifx \showCODEN    \undefined \def \showCODEN     #1{\unskip}     \fi
\ifx \showDOI      \undefined \def \showDOI       #1{#1}\fi
\ifx \showISBNx    \undefined \def \showISBNx     #1{\unskip}     \fi
\ifx \showISBNxiii \undefined \def \showISBNxiii  #1{\unskip}     \fi
\ifx \showISSN     \undefined \def \showISSN      #1{\unskip}     \fi
\ifx \showLCCN     \undefined \def \showLCCN      #1{\unskip}     \fi
\ifx \shownote     \undefined \def \shownote      #1{#1}          \fi
\ifx \showarticletitle \undefined \def \showarticletitle #1{#1}   \fi
\ifx \showURL      \undefined \def \showURL       {\relax}        \fi
% invisible to TeX
\providecommand\bibfield[2]{#2}
\providecommand\bibinfo[2]{#2}
\providecommand\natexlab[1]{#1}
\providecommand\showeprint[2][]{arXiv:#2}

\bibitem[Arnold et~al\mbox{.}(2021)]%
        {Arnold2021GenerativeMC}
\bibfield{author}{\bibinfo{person}{Kenneth~C Arnold}, \bibinfo{person}{April~M Volzer}, {and} \bibinfo{person}{Noah~G Madrid}.} \bibinfo{year}{2021}\natexlab{}.
\newblock \showarticletitle{Generative Models can Help Writers without Writing for Them}. In \bibinfo{booktitle}{\emph{Joint Proceedings of the IUI 2021 Workshops}}. \bibinfo{publisher}{CEUR-WS Team}, \bibinfo{address}{College Station, TX, USA}, \bibinfo{numpages}{8}~pages.
\newblock
\urldef\tempurl%
\url{http://ceur-ws.org/Vol-2903/}
\showURL{%
\tempurl}


\bibitem[Bassi et~al\mbox{.}(2023)]%
        {bassi2023end}
\bibfield{author}{\bibinfo{person}{Saksham Bassi}, \bibinfo{person}{Giulio Duregon}, \bibinfo{person}{Siddhartha Jalagam}, {and} \bibinfo{person}{David Roth}.} \bibinfo{year}{2023}\natexlab{}.
\newblock \bibinfo{title}{End-to-End Speech Recognition and Disfluency Removal with Acoustic Language Model Pretraining}.
\newblock
\newblock
\showeprint[arxiv]{2309.04516}~[eess.AS]
\urldef\tempurl%
\url{https://arxiv.org/abs/2309.04516}
\showURL{%
\tempurl}


\bibitem[Crystal(1995)]%
        {Crystal2005SpeakingOW}
\bibfield{author}{\bibinfo{person}{David Crystal}.} \bibinfo{year}{1995}\natexlab{}.
\newblock \showarticletitle{Speaking of writing and writing of speaking}.
\newblock \bibinfo{journal}{\emph{Longman Language Review}}  \bibinfo{volume}{1} (\bibinfo{year}{1995}), \bibinfo{pages}{5--8}.
\newblock


\bibitem[Dang et~al\mbox{.}(2022)]%
        {Dang2022BeyondTG}
\bibfield{author}{\bibinfo{person}{Hai Dang}, \bibinfo{person}{Karim Benharrak}, \bibinfo{person}{Florian Lehmann}, {and} \bibinfo{person}{Daniel Buschek}.} \bibinfo{year}{2022}\natexlab{}.
\newblock \showarticletitle{Beyond Text Generation: Supporting Writers with Continuous Automatic Text Summaries}. In \bibinfo{booktitle}{\emph{Proceedings of the 35th Annual ACM Symposium on User Interface Software and Technology}} (Bend, OR, USA) \emph{(\bibinfo{series}{UIST '22})}. \bibinfo{publisher}{Association for Computing Machinery}, \bibinfo{address}{New York, NY, USA}, Article \bibinfo{articleno}{98}, \bibinfo{numpages}{13}~pages.
\newblock
\showISBNx{9781450393201}
\urldef\tempurl%
\url{https://doi.org/10.1145/3526113.3545672}
\showDOI{\tempurl}


\bibitem[Fereday and Muir-Cochrane(2006)]%
        {fereday2006demonstrating}
\bibfield{author}{\bibinfo{person}{Jennifer Fereday} {and} \bibinfo{person}{Eimear Muir-Cochrane}.} \bibinfo{year}{2006}\natexlab{}.
\newblock \showarticletitle{Demonstrating Rigor Using Thematic Analysis: A Hybrid Approach of Inductive and Deductive Coding and Theme Development}.
\newblock \bibinfo{journal}{\emph{International Journal of Qualitative Methods}} \bibinfo{volume}{5}, \bibinfo{number}{1} (\bibinfo{year}{2006}), \bibinfo{pages}{80--92}.
\newblock
\urldef\tempurl%
\url{https://doi.org/10.1177/160940690600500107}
\showDOI{\tempurl}


\bibitem[Karat et~al\mbox{.}(1999)]%
        {Karat1999PatternsOE}
\bibfield{author}{\bibinfo{person}{Clare-Marie Karat}, \bibinfo{person}{Christine Halverson}, \bibinfo{person}{Daniel Horn}, {and} \bibinfo{person}{John Karat}.} \bibinfo{year}{1999}\natexlab{}.
\newblock \showarticletitle{Patterns of entry and correction in large vocabulary continuous speech recognition systems}. In \bibinfo{booktitle}{\emph{Proceedings of the SIGCHI Conference on Human Factors in Computing Systems}} (Pittsburgh, Pennsylvania, USA) \emph{(\bibinfo{series}{CHI '99})}. \bibinfo{publisher}{Association for Computing Machinery}, \bibinfo{address}{New York, NY, USA}, \bibinfo{pages}{568–575}.
\newblock
\showISBNx{0201485591}
\urldef\tempurl%
\url{https://doi.org/10.1145/302979.303160}
\showDOI{\tempurl}


\bibitem[Li et~al\mbox{.}(2021)]%
        {Li2021HierarchicalSF}
\bibfield{author}{\bibinfo{person}{Daniel Li}, \bibinfo{person}{Thomas Chen}, \bibinfo{person}{Albert Tung}, {and} \bibinfo{person}{Lydia~B Chilton}.} \bibinfo{year}{2021}\natexlab{}.
\newblock \showarticletitle{Hierarchical Summarization for Longform Spoken Dialog}. In \bibinfo{booktitle}{\emph{The 34th Annual ACM Symposium on User Interface Software and Technology}} (Virtual Event, USA) \emph{(\bibinfo{series}{UIST '21})}. \bibinfo{publisher}{Association for Computing Machinery}, \bibinfo{address}{New York, NY, USA}, \bibinfo{pages}{582–597}.
\newblock
\showISBNx{9781450386357}
\urldef\tempurl%
\url{https://doi.org/10.1145/3472749.3474771}
\showDOI{\tempurl}


\bibitem[Li et~al\mbox{.}(2023)]%
        {Li2023ImprovingAS}
\bibfield{author}{\bibinfo{person}{Daniel Li}, \bibinfo{person}{Thomas Chen}, \bibinfo{person}{Alec Zadikian}, \bibinfo{person}{Albert Tung}, {and} \bibinfo{person}{Lydia~B Chilton}.} \bibinfo{year}{2023}\natexlab{}.
\newblock \showarticletitle{Improving Automatic Summarization for Browsing Longform Spoken Dialog}. In \bibinfo{booktitle}{\emph{Proceedings of the 2023 CHI Conference on Human Factors in Computing Systems}} (Hamburg, Germany) \emph{(\bibinfo{series}{CHI '23})}. \bibinfo{publisher}{Association for Computing Machinery}, \bibinfo{address}{New York, NY, USA}, Article \bibinfo{articleno}{106}, \bibinfo{numpages}{20}~pages.
\newblock
\showISBNx{9781450394215}
\urldef\tempurl%
\url{https://doi.org/10.1145/3544548.3581339}
\showDOI{\tempurl}


\bibitem[Liao et~al\mbox{.}(2023)]%
        {Liao2020ImprovingRF}
\bibfield{author}{\bibinfo{person}{Junwei Liao}, \bibinfo{person}{Sefik Eskimez}, \bibinfo{person}{Liyang Lu}, \bibinfo{person}{Yu Shi}, \bibinfo{person}{Ming Gong}, \bibinfo{person}{Linjun Shou}, \bibinfo{person}{Hong Qu}, {and} \bibinfo{person}{Michael Zeng}.} \bibinfo{year}{2023}\natexlab{}.
\newblock \showarticletitle{Improving Readability for Automatic Speech Recognition Transcription}.
\newblock \bibinfo{journal}{\emph{ACM Trans. Asian Low-Resour. Lang. Inf. Process.}} \bibinfo{volume}{22}, \bibinfo{number}{5}, Article \bibinfo{articleno}{142} (\bibinfo{date}{May} \bibinfo{year}{2023}), \bibinfo{numpages}{23}~pages.
\newblock
\showISSN{2375-4699}
\urldef\tempurl%
\url{https://doi.org/10.1145/3557894}
\showDOI{\tempurl}


\bibitem[Lin et~al\mbox{.}(2024)]%
        {lin2024rambler}
\bibfield{author}{\bibinfo{person}{Susan Lin}, \bibinfo{person}{Jeremy Warner}, \bibinfo{person}{J.D. Zamfirescu-Pereira}, \bibinfo{person}{Matthew~G Lee}, \bibinfo{person}{Sauhard Jain}, \bibinfo{person}{Shanqing Cai}, \bibinfo{person}{Piyawat Lertvittayakumjorn}, \bibinfo{person}{Michael~Xuelin Huang}, \bibinfo{person}{Shumin Zhai}, \bibinfo{person}{Bjoern Hartmann}, {and} \bibinfo{person}{Can Liu}.} \bibinfo{year}{2024}\natexlab{}.
\newblock \showarticletitle{Rambler: Supporting Writing With Speech via LLM-Assisted Gist Manipulation}. In \bibinfo{booktitle}{\emph{Proceedings of the 2024 CHI Conference on Human Factors in Computing Systems}} (Honolulu, HI, USA) \emph{(\bibinfo{series}{CHI '24})}. \bibinfo{publisher}{Association for Computing Machinery}, \bibinfo{address}{New York, NY, USA}, Article \bibinfo{articleno}{1043}, \bibinfo{numpages}{19}~pages.
\newblock
\showISBNx{9798400703300}
\urldef\tempurl%
\url{https://doi.org/10.1145/3613904.3642217}
\showDOI{\tempurl}


\bibitem[Mehra et~al\mbox{.}(2023)]%
        {Mehra2023GistAV}
\bibfield{author}{\bibinfo{person}{Brinda Mehra}, \bibinfo{person}{Kejia Shen}, \bibinfo{person}{Hen~Chen Yen}, {and} \bibinfo{person}{Can Liu}.} \bibinfo{year}{2023}\natexlab{}.
\newblock \showarticletitle{Gist and Verbatim: Understanding Speech to Inform New Interfaces for Verbal Text Composition}. In \bibinfo{booktitle}{\emph{Proceedings of the 5th International Conference on Conversational User Interfaces}} (Eindhoven, Netherlands) \emph{(\bibinfo{series}{CUI '23})}. \bibinfo{publisher}{Association for Computing Machinery}, \bibinfo{address}{New York, NY, USA}, Article \bibinfo{articleno}{15}, \bibinfo{numpages}{11}~pages.
\newblock
\showISBNx{9798400700149}
\urldef\tempurl%
\url{https://doi.org/10.1145/3571884.3597134}
\showDOI{\tempurl}


\bibitem[Ruan et~al\mbox{.}(2018)]%
        {Ruan2018Comparing}
\bibfield{author}{\bibinfo{person}{Sherry Ruan}, \bibinfo{person}{Jacob~O. Wobbrock}, \bibinfo{person}{Kenny Liou}, \bibinfo{person}{Andrew Ng}, {and} \bibinfo{person}{James~A. Landay}.} \bibinfo{year}{2018}\natexlab{}.
\newblock \showarticletitle{Comparing Speech and Keyboard Text Entry for Short Messages in Two Languages on Touchscreen Phones}.
\newblock \bibinfo{journal}{\emph{Proc. ACM Interact. Mob. Wearable Ubiquitous Technol.}} \bibinfo{volume}{1}, \bibinfo{number}{4}, Article \bibinfo{articleno}{159} (\bibinfo{date}{Jan.} \bibinfo{year}{2018}), \bibinfo{numpages}{23}~pages.
\newblock
\urldef\tempurl%
\url{https://doi.org/10.1145/3161187}
\showDOI{\tempurl}


\bibitem[Tanaka et~al\mbox{.}(2018)]%
        {Tanaka2018NeuralEC}
\bibfield{author}{\bibinfo{person}{Tomohiro Tanaka}, \bibinfo{person}{Ryo Masumura}, \bibinfo{person}{Hirokazu Masataki}, {and} \bibinfo{person}{Yushi Aono}.} \bibinfo{year}{2018}\natexlab{}.
\newblock \showarticletitle{Neural Error Corrective Language Models for Automatic Speech Recognition}. In \bibinfo{booktitle}{\emph{INTERSPEECH}}. \bibinfo{publisher}{International Speech Communication Association}, \bibinfo{address}{Hyderabad, India}, \bibinfo{pages}{401--405}.
\newblock
\urldef\tempurl%
\url{https://doi.org/10.21437/Interspeech.2018-1430}
\showDOI{\tempurl}


\bibitem[Yang et~al\mbox{.}(2022)]%
        {Yang2022AIAA}
\bibfield{author}{\bibinfo{person}{Daijin Yang}, \bibinfo{person}{Yanpeng Zhou}, \bibinfo{person}{Zhiyuan Zhang}, \bibinfo{person}{{Toby Jia-Jun} Li}, {and} \bibinfo{person}{Ray LC}.} \bibinfo{year}{2022}\natexlab{}.
\newblock \showarticletitle{AI as an Active Writer: Interaction strategies with generated text in human-AI collaborative fiction writing}. In \bibinfo{booktitle}{\emph{Joint Proceedings of the IUI 2022 Workshops}} \emph{(\bibinfo{series}{CEUR Workshop Proceedings})}, \bibfield{editor}{\bibinfo{person}{Alison Smith-Renner} {and} \bibinfo{person}{Ofra Amir}} (Eds.). \bibinfo{publisher}{CEUR-WS Team}, \bibinfo{address}{Helsinki, Finland}, \bibinfo{pages}{56--65}.
\newblock
\urldef\tempurl%
\url{http://ceur-ws.org/Vol-3124/}
\showURL{%
\tempurl}


\end{thebibliography}

\appendix

\section{Appendix}
\begin{table*}[ht!]
    \caption{Participants Demographic Information, Experiences in Dictation and LLM Usages, Writing Practices}
    \label{tab:users}
    \centering
    \small
    \renewcommand{\arraystretch}{1.2}
    \begin{tabular}{|p{1.8cm}|p{2.3cm}|p{1.8cm}|p{2cm}|p{4cm}|p{3.5cm}|}
        \hline
        \textbf{Participant ID} & \textbf{Occupation} & \textbf{Dictation Usage Frequency} & \textbf{LLM Usage for English Writing Experience} & \textbf{Writing Tasks in Diary Study} & \textbf{Writing Contexts in Diary Study (Environments \& Devices)} \\ \hline
        P1* & Lecturer & A few times a year & Experienced user & - & - \\ \hline
        P2 & Ph.D. Student & A few times a year & Experienced user & Diaries; letter & On desk (computer); on bed (phone) \\ \hline
        P3 & Undergrad Student & A few times a month & Experienced user & Casual literary note; progress update report; feedbacks to teammates & Lying on bed (phone); walking dog (phone); riding vehicle (phone) \\ \hline
        P4 & Ph.D. Student & A few times a year & Frequent user & Prose; diaries & On bed (phone) \\ \hline
        P5 & Undergrad Student & A few times a month & Experienced user & Feedback to student as a part-time English tutor; new semester plan; notes about incidents in personal life & On desk (phone and computer); on bed (computer); in room that changing clothes (phone) \\ \hline
        P6 & Engineer & A few times a month & Occasional user & Documentation notes about emotion; notes about a milestone in life recently & In room (phone) \\ \hline
        P7* & Undergrad Student & Never & Frequent user & - & - \\ \hline
        P8 & Undergrad Student & A few times a month & Occasional user & Continuous updates for the novel that started previously & In workstation (phone); walking (phone) \\ \hline
        P9 & Ph.D. Student & A few times a month & Some experience with prompt engineering & Social media post; presentation script; more detailed outline for a research project & On bed (phone); office (computer) \\ \hline
        P10 & Blogger & A few times a year & Occasional user & Blog posts about reflection towards certain social phenomenon & Walking (phone) and sitting at home (digital tablet); home (phone); home (computer) \\ \hline
        P11* & Grad Student & A few times a week & Occasional user & - & - \\ \hline
        P12 & Grad Student & Never & Occasional user & Screenplays; self-critique report to other screenplays & Home (computer) \\ \hline
        P13 & UX Designers & A few times a year & Occasional user & Fictional story based on the characters in a famous game; casual emotion notes; notes about a milestone in life recently & Home (computer) \\ \hline
        P14 & UX Designers & A few times a month & Occasional user & Casual emotion notes; thoughts and comments about a movie & Home (computer); walking (phone) \\ \hline
        P15 & Professor & Everyday & Experienced user & Potential blog posts (reflection and thoughts about some social phenomenon) & Walking (phone); home (phone); office (computer) \\ \hline
    \end{tabular}
    \par\vspace{0.5em} % Adds space before the notation
    \parbox{\textwidth}{\textit{Note: Cells marked with * represent participants who were absent in the diary study.}}
\end{table*}









\end{document}